\documentclass[twocolumn]{aastex631}
\usepackage{graphicx,amssymb,amsmath,amsfonts,multirow}
\usepackage{subfigure,epstopdf,threeparttable}
\usepackage{CJK}

\begin{document}
\begin{CJK*}{UTF8}{gbsn}
\title{The Formation of Ultra-short-period Planets under the Influence of the Nearby Planetary Companions}
 
\correspondingauthor{Su Wang}
\email{wangsu@pmo.ac.cn}
\correspondingauthor{Jianghui Ji}
\email{jijh@pmo.ac.cn}

\author[0009-0001-0885-4903]{Jiajun Zhu (朱家骏)}
\affiliation{CAS Key Laboratory of Planetary Sciences, Purple Mountain Observatory, Chinese Academy of Sciences, Nanjing 210023, China}
\affiliation{School of Astronomy and Space Science, University of Science and Technology of China, Hefei 230026, China}

\author[0000-0002-4859-259X]{Su Wang (王素)}
\affiliation{CAS Key Laboratory of Planetary Sciences, Purple Mountain Observatory, Chinese Academy of Sciences, Nanjing 210023, China}
\affiliation{School of Astronomy and Space Science, University of Science and Technology of China, Hefei 230026, China}
\affiliation{CAS Center for Excellence in Comparative Planetology, Hefei 230026, China}

\author[0000-0002-9260-1537]{Jianghui Ji (季江徽)}
\affiliation{CAS Key Laboratory of Planetary Sciences, Purple Mountain Observatory, Chinese Academy of Sciences, Nanjing 210023, China}
\affiliation{School of Astronomy and Space Science, University of Science and Technology of China, Hefei 230026, China}
\affiliation{CAS Center for Excellence in Comparative Planetology, Hefei 230026, China}

\author[0000-0001-8162-3485]{Yao Dong (董瑶)}
\affiliation{CAS Key Laboratory of Planetary Sciences, Purple Mountain Observatory, Chinese Academy of Sciences, Nanjing 210023, China}
\affiliation{CAS Center for Excellence in Comparative Planetology, Hefei 230026, China}

\begin{abstract}
  Ultra-short-period (USP) planets, defined as those with orbital periods shorter than 1 day, provide valuable insights into planetary evolution under strong stellar tidal interactions. In this work, we investigate the formation of USP planets in two-planet systems consisting of an inner terrestrial planet accompanied by an outer hot Jupiter (HJ). Our simulation results show USP planets can form through a process driven by secular perturbations from the outer companion, which induce eccentricity excitation, tidal dissipation, and subsequent orbital decay of the inner planet. The probability of USP formation is governed by key factors, including the mass ratio between two planets, their orbital eccentricities, and the tidal dissipation process. 6.7\% of our simulations form USP planets, and USP planets form most efficiently when the mass ratio is around 4 $M_{\oplus}{\rm /}M_{\rm J}$, with the inner planet less than 8 $M_{\oplus}$. Furthermore, the eccentricity of the outer HJ plays a crucial role-moderate eccentricities ($e_{\rm outer}<0.1$) favor USP formation, whereas higher eccentricities ($e_{\rm outer}>0.1$) enhance the likelihood of orbital instability, often resulting in a lonely HJ. USP planets form more efficiently when the tidal dissipation function of the inner planet is comparable to the values estimated for terrestrial planets in the solar system. Comparison with observed planetary systems reveals that systems with large mass ratios or nearly circular outer planets tend to produce short-period (SP) planets instead of USP planets. Our findings offer a potential explanation for the most commonly observed system architectures, which predominantly feature either an HJ with an inner SP planet or a lonely HJ.
\end{abstract}

\section{INTRODUCTION}
\begin{figure*}
    \centering
    \includegraphics[width=1\linewidth]{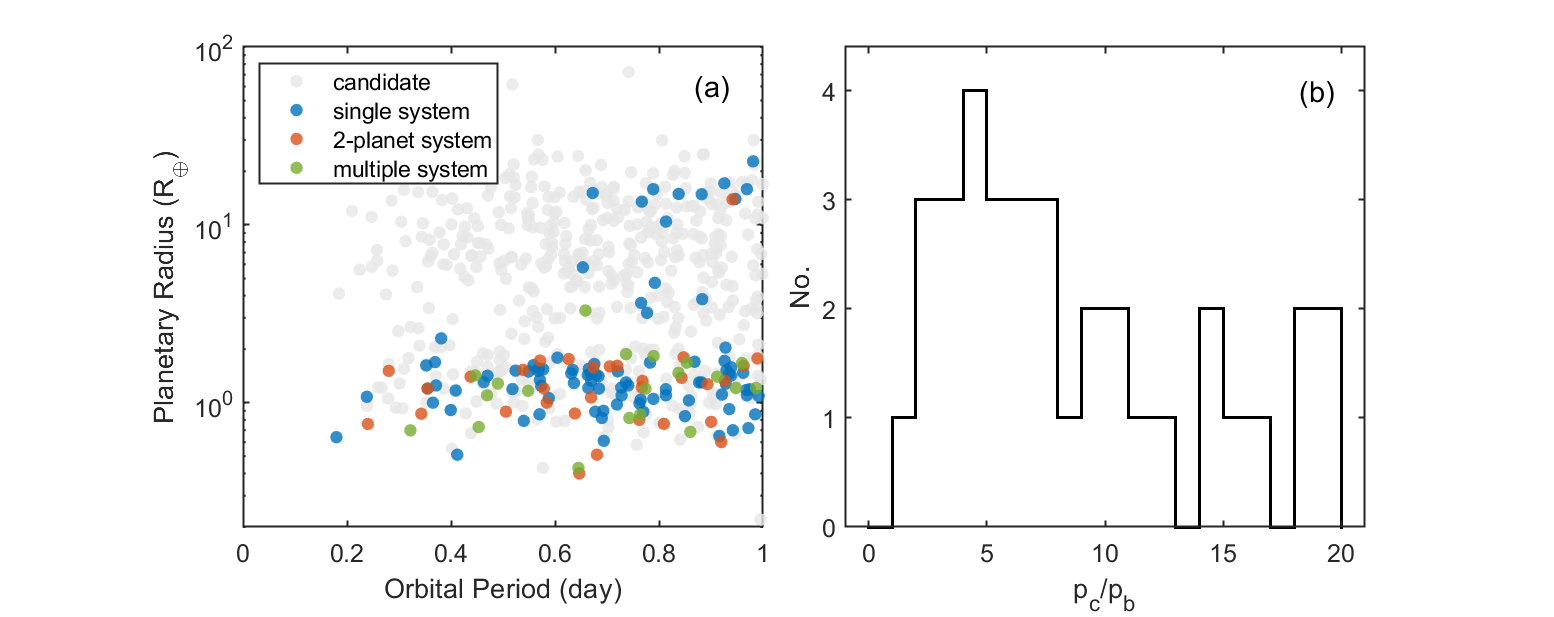}
    \caption{(a) The distribution of orbital periods and planetary radii for USP planets. The blue, red, and green dots represent USP planets confirmed in systems with one, two, and more than two planets, respectively. The gray dots indicate USP candidates that have not been confirmed yet. (b) The period ratios between confirmed USP planets and their adjacent planets in multiple systems, $p_{b}$ denotes the period of the USP planet, and $p_{c}$ represents the orbital period of its nearest outer neighbor.}
    \label{fig:1}
\end{figure*}

As of 2024 December, approximately 141 ultra-short-period (USP) planets with orbital periods of less than 1 day have been detected, accounting for 2.44\% of all confirmed exoplanets \citep{2014ApJ...787...47S}. In addition to these confirmed USP planets, about 502 USP candidates are awaiting confirmation. Figure \ref{fig:1} displays the distribution of masses and orbital periods for confirmed USP planets and candidates \citep{2014ApJ...787...47S,2020DPS....5230306A, 2021ApJ...919...26U}. Among the confirmed USP planets, K2-137 b, which orbits around a 0.463 $M_{\odot}$ M-dwarf star, has the shortest orbital period, about 0.18 days, and its mass is about 0.89 $M_{\oplus}$ \citep{2018MNRAS.474.5523S, 2021PSJ.....2..152A}. Considering additional candidate planets, KOI 1843.03 is estimated to have an orbital period about 4 minutes shorter than that of K2-137 b \citep{2013ApJ...773L..15R}. As shown in Figure \ref{fig:1}, approximately 90\% of the confirmed USP planets have radii smaller than 2 $R_{\oplus}$ corresponding to a mass of $\sim$10 $M_\oplus$ based on the estimation of the mass-radius relationship $M_{p}=M_{\oplus}(R_{p}/R_{\oplus})^{3.7}$ \citep{2006Icar..181..545V}. The remaining 10\% consists of larger planets with radii ranging from 2 to 20 $R_\oplus$. About one-third of the USP planets are found in multiple planetary systems, as shown in the red and green dots in Figure \ref{fig:1}, where 31 USP planets are found in two-planet systems, 11 USP planets in three-planet systems, six in four-planet systems, two in five-planet systems, and one in a six-planet system. In particular, among the multiple planetary systems with USP planets, only one system contains two USP planets, Kepler-70, which includes two small planets in an extremely compact configuration \citep{2011Natur.480..496C}. Although most USP candidates are in single-planet systems, it remains possible that additional planets, especially distant companions, could be found in these systems in the future. As shown in Panel (b) of Figure \ref{fig:1}, 91\% of the period ratios between the confirmed USP planets and their adjacent planet exceed 3.0, and 47\% have period ratios greater than 10.0, implying that most USP planets are far from their planetary neighbors. From the orbital characteristics of USP planets in multiple planetary systems, we can deduce that they are located extremely close to their host stars, resulting in strong interactions with the star, while being distant from other planets in the system. These features of orbital configurations provide clues about the formation and evolutionary history of USP planets.

Several likely scenarios have been proposed for the formation of USP planets. One possible explanation is that the USP planets are the leftover dense cores after the evaporation of gas giants. The stripping of their H/He atmosphere could be caused by photoevaporation \citep{2013ApJ...769L...9B,2013ApJ...776....2L,2018MNRAS.477..808L} or by the core-powered mass loss mechanism \citep{2012ApJ...753...66I,2016ApJ...825...29G}. Terrestrial USP planets with $R<2\ R_{\oplus}$ may originate from sub-Neptunes with the stripping of the H/He atmosphere, which could explain the observed absence of USP sub-Neptunes \citep{2017MNRAS.472..245L}. The Roche lobe overflow of hot Jupiters (HJs) can also lead to the formation of USP planets with $R>5\ R_{\oplus}$. However, if the USP planets originate from HJs \citep{2014ApJ...793L...3V}, the metallicities and masses of their host stars should be higher than those of the Sun, which differs from the average properties of the host stars of the known USP planets \citep{2017AJ....154...60W}. An eccentricity exceeding 0.8 of the innermost planet due to secular chaos with the outer companion planets may also be responsible for the formation of USP planets. In this scenario, the tidal effects caused by the central star would eventually circularize the orbit of the planet and make it closer to the central star \citep{2007MNRAS.382.1768M,2008IAUS..249..285Z}. This high-eccentricity migration mechanism requires a companion with a minimum mass of 10 $M_{\oplus}$ and an orbital period exceeding 10 days \citep{2019AJ....157..180P}. Alternatively, a low-eccentricity migration mechanism could also lead to the formation of a USP planet. In multiple planetary systems, modest eccentricities, approximately 0.1-0.2, can drive planets into USP planets due to enhanced eccentricity excitation and increased mutual inclination caused by apsidal and nodal precession resonances \citep{2019MNRAS.488.3568P}. In this scenario, systems require additional planets to provide an angular momentum deficit (AMD) to sustain the tidal decay process, especially for the systems with more than three terrestrial planets are more likely to form USP planets, with the mass of companion planets in the range of 3-20 $M_{\oplus}$.

Based on the previously mentioned formation scenarios and the distribution of period ratios between the USP planets and their nearby planetary companions in multiple planetary systems, eccentricity excitation is considered an important process in the formation of USP planets. In multiple planetary systems, several possible processes may contribute to the eccentricity excitation of the USP planets.

\begin{table*}
	\caption{Detailed Information on the Observed Eight Systems that Hold Hot Jupiters and Inner Low-mass Companions}
\begin{center}
	\begin{tabular}{ccccccc}
		\hline \hline
		\multicolumn{1}{l}{} &  Planet Name& Mass  & $M_{i}/M_{o}$    & Eccentricity  & Orbital period  & $P_{o}/P_{i}$\\
		&&&($M_{\oplus}/M_{\rm J}$)&&(day)& \\
		\hline
		\multirow{4}{*}{WASP-47} & e      & 6.77 $M_{\oplus}$ & 5.92 &  0  &  0.78961  & 5.27 \\
		& b      & 1.144 $M_{\rm J}$ & $\ldots$ & 0.00060  & 4.1591510  & 2.17  \\
		& d      & 15.5 $M_{\oplus}$ &$\ldots$& 0.0010  & 9.030501  & 65.3   \\
		& c      & 1.253 $M_{\rm J}$ &$\ldots$& 0.264  & 589.57  & $\ldots$  \\
		\hline
		\multirow{2}{*}{WASP-132} & c      & 6.26 $M_{\oplus}$ & 14.6 & 0.13  & 1.011534  & 7.05 \\
		& b      & 0.428 $M_{\rm J}$ & $\ldots$ & 0.070  & 7.133514  & $\ldots$  \\
		\hline
		\multirow{2}{*}{WASP-84} & c      & 15.2 $M_{\oplus}$ &   22.0 &$\ldots$& 1.4468849  & 5.89 \\
		& b      & 0.692 $M_{\rm J}$ &   $\ldots$&$\ldots$& 8.52349648  &  $\ldots$ \\
		\hline
		\multirow{2}{*}{Kepler-975} & b      & $\sim$4.27 $M_{\oplus}$ & 10.3 & $\ldots$ & 1.970342457  & 2.57 \\
		& c      & $\sim$0.414 $M_{\rm J}$ &  $\ldots$&$\ldots$ & 5.05821991  & $\ldots$  \\
		\hline
		\multirow{2}{*}{TOI-1408} & c      & 7.6 $M_{\oplus}$ & 4.06 & 0.1353  & 2.1664  & 2.04 \\
		& b      & 1.87 $M_{\rm J}$ &$\ldots$& 0.0023  & 4.42587  & $\ldots$  \\
		\hline
		\multirow{2}{*}{Kepler-730} & c      & $\sim$5.31 $M_{\oplus}$ & 9.55 & $\ldots$  & 2.851883380  & 2.28 \\
		& b      & $\sim$0.556 $M_{\rm J}$ &  $\ldots$&$\ldots$ & 6.491682808  & $\ldots$  \\
		\hline
		\multirow{2}{*}{TOI-1130} & b      & 19.8 $M_{\oplus}$ & 20.3 & 0.22  & 4.066499  & 2.05 \\
		& c      & 0.9740 $M_{\rm J}$ &$\ldots$& 0.047  & 8.350381  & $\ldots$  \\
		\hline
		\multirow{2}{*}{TOI-5398} & c      & 11.8 $M_{\oplus}$ & 63.8 & $<$0.14  & 4.77271  & 2.22 \\
		& b      & 0.185 $M_{\rm J}$ &$\ldots$& $<$0.13  & 10.590547  & $\ldots$  \\
		\hline
	\end{tabular}
	\label{obs}
\end{center}
\begin{tablenotes}
	\item \textbf{Note.} The $M_{i}/M_{o}$ shows the mass of the inner low-mass companion divided by the mass of the hot Jupiter. And $P_{o}/P_{i}$ means the period ratio between two adjacent planets.
\end{tablenotes}
\end{table*}

Among the confirmed multiple planetary systems with USP planets, some contain planet pairs that are in or near mean motion resonances (MMRs). Kepler-80 contains one USP planet and five outer companions, with the outer companions residing in or near 3:2, 3:2, 4:3, and 3:2 MMRs, respectively \citep{2016AJ....152..105M}. Kepler-32 has five confirmed planets, in addition to one USP planet, with three of the other four planets in or near 2:1 and 3:2 MMRs \citep{2012ApJ...750..114F}.  In the WASP-47 system, the USP planet WASP-47 e is accompanied by a HJ in a near 2:1 MMR configuration with a more distant planetary companion \citep{2015ApJ...812L..18B, 2023A&A...673A..42N}. On the one hand, orbital migration is an important process in the formation of planet pairs within or near MMRs configuration \citep{2012ApJ...753..170W,2014ApJ...795...85W}. On the other hand, theoretical estimates of the isolation mass \citep{1987Icar...69..249L,1998Icar..131..171K,2000Icar..143...15K,2020apfs.book.....A} in the observed short orbital periods of USP planets are significantly smaller than the mass predicted by the mass-radius relationship. This suggests that USP planets, especially those confirmed in the multiple planetary systems, are more likely to form via orbital migration from the outer regions of the systems. Once planets are captured into MMRs, their eccentricities can be excited \citep{2014ApJ...795...85W}. With an outer massive planetary companion, the eccentricity of the inner planet can be excited \citep{2007MNRAS.382.1768M}. \cite{2025arXiv250300872W} suggest that WASP-47 was formed under the MMRs trapping, outer massive planet perturbation, and the tidal effects, which are three major processes leading to the formation of USP planets. Theoretical predictions for the positions of the three planets differ from observations by no more than 4\%. However, the formation of USP planets through this scenario is also influenced by the mass ratio, the semimajor axis ratio between the USP and the outer planetary companions, and the eccentricity-damping process of the tidal effect. Through confirmed planetary systems, seven systems contain an HJ ($M > 100\ M_\oplus$) with an inner low-mass companion: TOI-1130, WASP-132, Kepler-730, Kepler-975, TOI-1408, WASP-47, and WASP-84 \citep{2015ApJ...812L..18B,2016ApJ...822...86M,2019ApJ...870L..17C,2020ApJ...892L...7H,2022AJ....164...13H, 2023MNRAS.525L..43M, 2024ApJ...971L..28K}. Additionally, TOI-5398, which hosts a Saturn-mass planet for nearly 10 days with an inner companion, exhibits a configuration similar to the inner two planets in the WASP-47 system \citep{2024A&A...682A.129M, 2024A&A...684L..17M}. Table \ref{obs} shows the detailed information on these systems. Among them, five systems have the inner planetary pairs near 2:1 MMR,  WASP-47 and WASP-132 (the orbital period of WASP-132 c is 1.01, which can be classified as a USP planet) have a USP planet in the system. The fraction of configurations with HJ and inner USP companion among the total USP population is quite low, about 1.4\% .

In this work, we explore the formation of USP planets in multiplanet systems, with a particular focus on the influence of nearby massive companions and the tidal effects exerted by the central star. Our goal is to determine how the presence of an outer massive planet impacts USP planetary formation. Section 2 outlines the models employed in this study, while Section 3 presents the major simulation results. Finally, we summarize our conclusions in Section 4.

\section{MODELS}
\subsection{Dynamical Effects}

To investigate the formation of USP planets under the influence of nearby planetary companions, we assume a system with a Sun-like star, as most stars hosting USP planets are of this type \citep{2017AJ....154...60W,2020ChA&A..44..283H}, surrounded by two planets, an inner super-Earth with a mass of less than 10 $M_\oplus$ (planet b) and an outer planet (planet c) with a mass no less than 0.5 $M_{\rm J}$. We mainly focus on the dynamical evolution after the dissipation of the protoplanetary disk, where the dominant forces acting on the planets are the gravitational interactions with other bodies and the tidal effects caused by the central star, as

\begin{equation}
\begin{aligned}
\frac{d}{dt}\mathbf{V}_i ={} & -\frac{G\,(M_*+M_i)}{r_i^2}\left(\frac{\mathbf{r}_i}{r_i}\right) \\
& + \sum_{j\neq i}^N G\,M_j \left[\frac{(\mathbf{r}_j-\mathbf{r}_i)}{|\mathbf{r}_j-\mathbf{r}_i|^3} - \frac{\mathbf{r}_j}{r_j^3}\right] \\
& + \mathbf{f}_{\rm tidal}\,.
\end{aligned}
\end{equation}

where $\textbf{\textit{V}}_i$, $\textbf{\textit{r}}_i$, $r_i$, and $M_*$ mean the velocity and position vectors of the planet, the distance from the central star, and the mass of the central star, respectively. $\textbf{\textit{f}}_{\rm tidal}$ represents the tidal force caused by the central star \citep{2002ApJ...573..829M} as shown
\begin{equation}
    \begin{aligned}
        \textbf{\textit{f}}_{\rm tidal} =& - \left( \dfrac{3nk_{Lp}}{Q_{p}} \right) \left( \dfrac{M_{*}}{M_{p}} \right) \left( \dfrac{R_{p}}{a} \right)^{5} \left( \dfrac{a}{r} \right)^{8} \\
        & \times \left[ 3\left( \pmb{\hat{r}} \cdot \pmb{\dot{r}} \right) +\left(\pmb{\hat{r}} \times \pmb{\dot{r}}-r{\pmb{\Omega}} \right)\times \pmb{\hat{r}}\right],
    \end{aligned}
    \label{tidal}
\end{equation}
where $R_{p}$, $n$ and $\pmb{\textit{r}}$ are the radius, mean motion of planets, and the relative position vector from the star to the planet, respectively. $M_{p}$ and $M_{*}$ are the masses of the planets and the host star. For terrestrial planets with $R_{p} < 2\ R_{\oplus}$, there is the mass-radius relationship $M_{p}=M_{\oplus}(R_{p}/R_{\oplus})^{3.7}$ \citep{2006Icar..181..545V}. $\pmb{\Omega}$ is the spin rate of the planets, with the value equal to $n$ due to the tidal locking  \citep{1989A&A...223..112Z,2002A&A...386..222W,2008EAS....29....1M}. $k_{Lp}$ and $Q_{p}$ represent the Love number and the tidal dissipation function of the planets, both of which are related to the internal properties of the planets. Here, we assume the Love number $k_{Lp}$ to be a constant 1.5. Under tidal effects, planetary eccentricity is expected to be circularized on a timescale given by \citep{2008IAUS..249..285Z}
\begin{equation}
\tau_{\rm circ}=\dfrac{4Q_{p}}{63n}\dfrac{M_{p}}{M_{*}}\left(\dfrac{a}{R_{p}}\right)^{5}.
    \label{circ}
\end{equation}
The tidal dissipation function $Q_{p}$ of planets in the solar system has been measured or constrained. For terrestrial planets, $Q_p$ typically ranges from $10^{1}$ to $10^{3}$ \citep{1975CeMec..12...61S,1978VA.....22..133S,1982LPI....13..190D,1999ssd..book.....M}, while for giant planets, it falls within $10^{4} \lesssim Q_{p} \lesssim 10^{6}$ \citep{1981Icar...47....1Y,1990Icar...85..394T,2008Icar..193..213M,2008Icar..193..267Z}. Since the tidal effects caused by the planets on the star or between the planets are significantly weaker than those induced by the host star on the planets, we neglect these effects in this work.

\subsection{Initial Settings of the Numerical Simulations}

The initial planetary configuration is obtained through an orbital migration process when the protoplanetary disk still exists. The MMR configurations are related to the density profile of the protoplanetary disk, the migration speed and direction of the planets \citep{2014ApJ...795...85W,2017AJ....154..236W}. Based on estimates of migration timescales, the inner planet undergoes type I migration, which is much faster than the type II migration experienced by the outer massive planet. As a result, the inner planet migrates to the location near the inner edge of the protoplanetary disk, while the outer massive planet still continues its inward migration. This process typically results in a period ratio greater than 2.0, which leads to a high opportunity to be trapped in a 2:1 MMR. Numerical simulations of a two-planet system, consisting of a terrestrial planet and a Jupiter-like planet embedded in a protoplanetary disk similar to a minimum-mass solar nebular \citep{1981PThPS..70...35H} with a depletion timescale of $10^6$ yr, indicate that most of the planet pairs in these systems can be trapped into 2:1 MMRs. In this work, we choose the orbital parameters of the planets derived from numerical simulations after the orbital migration process in a two-planet system as the initial conditions, as shown in Table \ref{mig}. The eccentricity of the inner terrestrial planet can be excited in the range [0.152, 0.263], while the eccentricity of the outer giant remains between [0.044, 0.055], significantly lower than that of the inner planet. Since the mass ratio between the two planets, the eccentricity, and the $Q_p$ of the planets are the three key factors that may affect the formation of USP planets, we conduct three groups of simulations to explore their effects.

\begin{table*}
\caption{Orbital Parameters of Two Planets in the Systems Obtained through Orbital Migration Process}
\begin{center}
    \begin{tabular}{ccccc}
    \hline \hline
      Planet & Mass & Orbital Period  & Semimajor Axis  & Eccentricity \\
      &&(day)&(au)&\\
    \hline
     b   & 4 $M_{\oplus}$  & 2.5  & 0.03607  & 0.152-0.263 \\
     c   & 1 $M_{\rm J}$  & 5.0  & 0.05725  & 0.0439-0.0546  \\
    \hline
    \end{tabular}
\label{mig}
\end{center}
\end{table*}

{\bf \textit{Group 1.}} Systems with various planetary mass ratios. In order to investigate the influence of the masses of the planets on the final planetary configuration, we choose $M_{b}=$1, 2, 4, and 8 $M_{\oplus}$ \citep{2021ApJ...919...26U} and $M_{c}=$0.5, 1, and 2 $M_{\rm J}$ \citep{2018ARA&A..56..175D}. This results in 12 pairs of planets with different masses in this group. The tidal dissipation function $Q_p$ is fixed at $10^{2}$ for the terrestrial planet and $10^{4}$ for the massive outer planet \citep{2013ApJ...774...52L}.

{\bf \textit{Group 2.}} Systems with different eccentricities of the planets. In this group, we select three possible values of the eccentricities of each planet. When the eccentricity of the inner planet is fixed at 0.2, the eccentricity of the outer planet varies among (0, 0.05, 0.10). In the same way, when the eccentricity of the outer planet is fixed at 0.05, the eccentricity of the inner planet can be chosen among (0.1, 0.2, 0.3). Thus, there are five possible settings for the planetary pair in one system. Here, $M_{b}$ and $M_{c}$ are fixed at 2 $M_{\oplus}$ and 0.5 $M_{\rm J}$, respectively.

{\bf \textit{Group 3.}} Systems with different planetary tidal dissipation functions $Q_p$. According to $Q_p$ of the planets in the Solar system, we assume that $Q_{pb}$ of the inner planet is fixed at $10^{2}$ when the $Q_{pc}$ of the outer planet varies among ($10^{4}$, $10^{5}$, $10^{6}$). Similarly, when $Q_{pc}$ is fixed at $10^{4}$, $Q_{pb}$ can choose among ($10^{0}$, $10^{1}$, $10^{2}$, $10^{3}$).
In this group, the eccentricities and masses of the planets are kept constant.

We perform 1050 simulations in total, 50 runs for each pair of planets with different initial longitudes of the ascending node $\Omega$, argument of periapsis $\omega$, and mean anomaly $M$, while other parameters are chosen as shown in Table \ref{group}. The orbital angles of planets in the 50 runs are chosen from 50 points along orbits that are in MMR. The resonant angles between the two planets are librating. In the simulation, two planets are coplanar. We use the \textit{N}-body integrator MERCURY6 \citep{1999MNRAS.304..793C} to study the dynamical evolution process of the systems. The integration time step is chosen to be less than 1/50 of the orbital period of the inner planets with an integration accuracy parameter of $10^{-16}$.

\begin{table*}
\caption{Initial Planetary Parameters in the Three Groups}
\begin{center}
    \begin{tabular}{ccccccc}
    \hline \hline
    \multicolumn{1}{l}{} &  Planet & Mass      & Orbital Period  & Semimajor Axis  & Eccentricity & $Q_{p}$\\
    &&&(day)&(au)&&\\
    \hline
    \multirow{2}{*}{Group 1} & b      & 1/2/4/8 $M_{\oplus}$ & 2.5  & 0.03607  & 0.2 & 10$^{2}$ \\
    & c      & 0.5/1/2 $M_{\rm J}$ & 5.0  & 0.05725  & 0.05  & 10$^{4}$       \\
    \hline
    \multirow{2}{*}{Group 2} & b      & 2 $M_{\oplus}$ & 2.5  & 0.03607  & (0.1/0.2/0.3), 0.2 & 10$^{2}$ \\
    & c      & 0.5 $M_{\rm J}$ & 5.0  & 0.05725  & 0.05, (0/0.05/0.10)  & 10$^{4}$       \\
    \hline
    \multirow{2}{*}{Group 3} & b      & 2 $M_{\oplus}$ & 2.5  & 0.03607  & 0.2 & (10$^{0}$/10$^{1}$/10$^{2}$/10$^{3}$), 10$^{2}$ \\
    & c      & 0.5 $M_{\rm J}$ & 5.0  & 0.05725  & 0.05  & 10$^{4}$, (10$^{4}$/10$^{5}$/10$^{6}$)       \\
    \hline
    \end{tabular}
\label{group}
\end{center}
\end{table*}

\section{NUMERICAL SIMULATION RESULTS}

\subsection{Three Types of Typical Results}
\begin{figure*}
    \centering
    \includegraphics[width=1.0\linewidth]{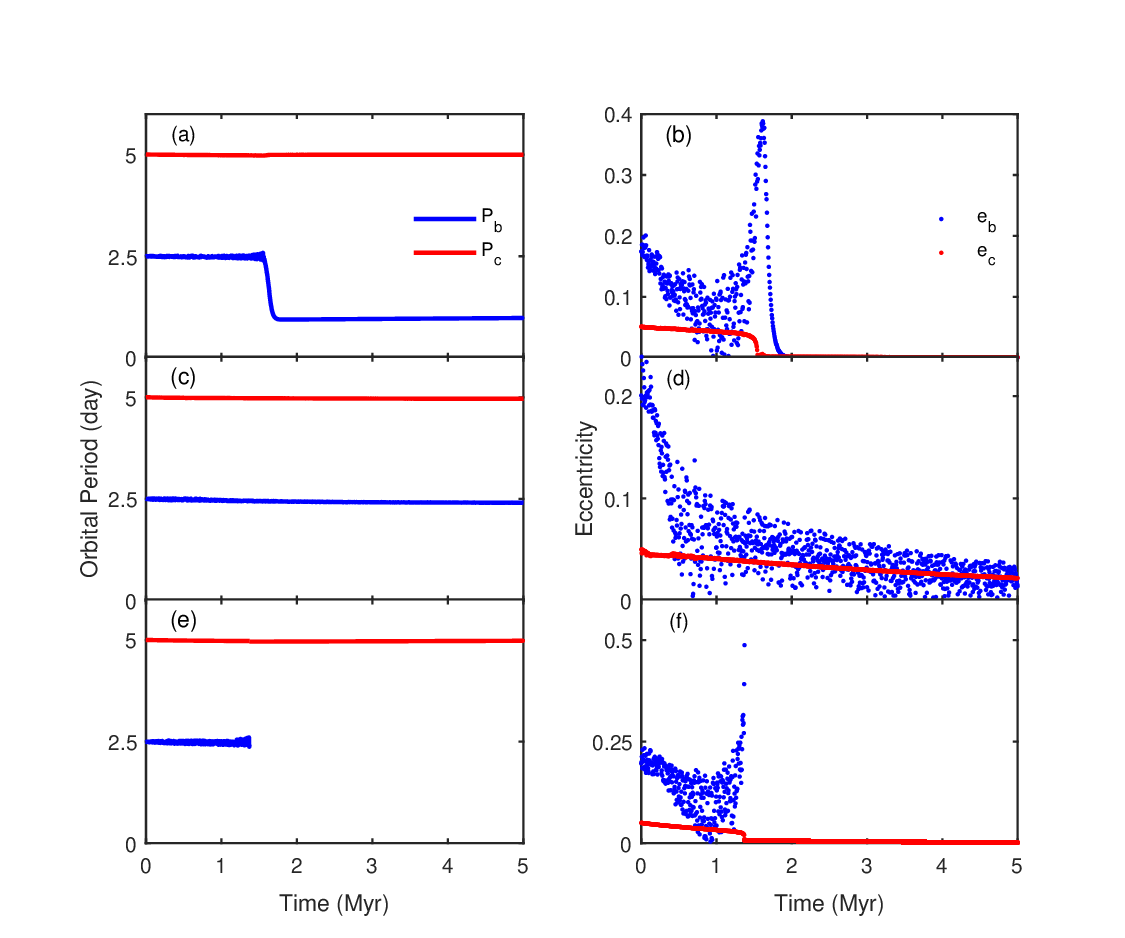}
    \caption{Evolution of three typical cases. Panels (a) and (b) illustrate the evolution of orbital periods and eccentricities in Case 1, where a USP planet forms. Panels (c) and (d) show the evolution of planets in Case 2, resulting in the formation of an SP planet. Panels (e) and (f) depict the evolution of a typical case with instability. The blue and red lines and dots represent the evolution of inner planet b and outer planet c, respectively.}
    \label{fig:3}
\end{figure*}

Based on our numerical simulations, three distinct planetary configurations emerge by the end of the integration. The first type is the configuration with HJ and inner USP planet formed in the system, the second type features an HJ accompanied by an inner short-period planets (SP planets, the planets with orbital period between [1, 10] days) form, and the third type results in the ejection of the terrestrial planet, leaving behind one lonely HJ in the system. Figure \ref{fig:3} shows three typical results with the USP planet (Case 1, the upper panels), SP planets (Case 2, the middle panels), and lonely HJ (Case 3, the lower panels) formed in the final planetary configuration. The initial orbital periods, eccentricities, and the $Q$-values of two planets are chosen as the same in group 1 as shown in Table \ref{group}. The major differences among Cases 1, 2, and 3 are the masses of planets b and c. In Case 1, the masses of two planets are 2 $M_\oplus$ and 0.5 $M_{\rm J}$, in Case 2, their masses are 4 $M_\oplus$ and 0.5 $M_{\rm J}$, and in Case 3, they are 1 $M_\oplus$ and 1 $M_{\rm J}$.

In Case 1, at the beginning of the simulation, the eccentricity of planet b gradually decreases due to the tidal dissipation combined with the minor excitation until $\sim$ 1.5 Myr. Meanwhile, the eccentricity of the outer giant steadily decreases to nearly 0. Around 1.5 Myr, the MMR configuration established through orbital migration is disrupted. The eccentricity of planet b was excited to nearly 0.4, which may be due to the long MMR process. However, due to the efficient tidal effect caused by the central star, the eccentricity of planet b can be damped down quickly, combined with a significant inward migration of planet b, as shown in Panel (a) of Figure \ref{fig:3}, ultimately resulting in the formation of a USP in the system. After the MMR disruption, the orbital period ratio expands from 2.0 to 5.31. By the end of the simulation, the orbital period of planet b decreases to about 0.97 days. The evolution process is similar to that of the WASP-47 system, which hosts three planets \citep{2025arXiv250300872W}. Similarly, the WASP-132 system, which hosts a $1.85~R_\oplus$ planet with an orbital period of approximately 1.01 days and an outer massive planet of $0.897~R_J$ at about 7.13 days \citep{2022AJ....164...13H}, may have undergone a similar evolution process.

In Case 2, the outcome differs significantly from Case 1. In this case, the eccentricity of planet b continuously decreases throughout the simulation. Although small amplitudes of eccentricity excitations occur, the overall trend of the eccentricity of planet b remains a steady decline. The MMR is disrupted before 1 Myr, much shorter than Case 1, leading to no significant excitation of eccentricity in Case 2. The eccentricity of the inner planet is expected to be excited to a maximum of approximately 0.28 according to the estimation \citep{2007MNRAS.382.1768M}, and the tidal effect makes the eccentricity damp to below 0.05 quickly. As a result, the orbital period ratio between the two planets remains close to 2.05, maintaining a configuration near the 2:1 MMR. Consequently, planet b undergoes only a slight inward migration from its initial location while preserving the 2:1 MMR, resulting in a stable SP planet at the end of the simulation. The obtained planetary configuration closely resembles that of Kepler-730, TOI-1130, and TOI-5398, which contain a short-period low-mass inner planet and a massive outer planet \citep{2019ApJ...870L..17C,2020ApJ...892L...7H,2024A&A...682A.129M, 2024A&A...684L..17M}. The orbital period ratios between adjacent planets in these three systems are 2.28, 2.05, and 2.22, respectively. The period ratios are all slightly above 2.0, which is consistent with the results in Case 2 implying a similar formation path of these systems: the orbit of the inner planet is gradually circularized due to the tidal effects caused by the central star combined with the increase of the separation between the two planets, causing the planet pair moves slightly away from their original resonant position.

In addition to results similar to Cases 1 and 2, orbital crossing occurred in 96 out of all 1050 runs. In these runs, the inner low-mass planet either collides with the outer Jupiter-like planet, is scattered out of the system, or merges with the central star, leaving behind a lonely HJ at the end of the simulation. Panels (e) and (f) show a typical case in which an orbital crossing occurred. The inner planet collides with the outer HJ at about 1.4 Myr. The eccentricity of the inner planet can be excited to greater than 0.5, which is the main reason for the instability. Before the instability, the system evolves similarly to Case 1.

\subsection{The Ultra-short-period Planets Formation by Different Effects}
We carry out 1050 runs to investigate the formation of short-period planets influenced by the planetary masses, the eccentricities of the planets, and the tidal effects induced by the central star.

\subsubsection{Group 1: Planetary Masses}
\begin{figure*}
    \centering
    \includegraphics[width=1\linewidth]{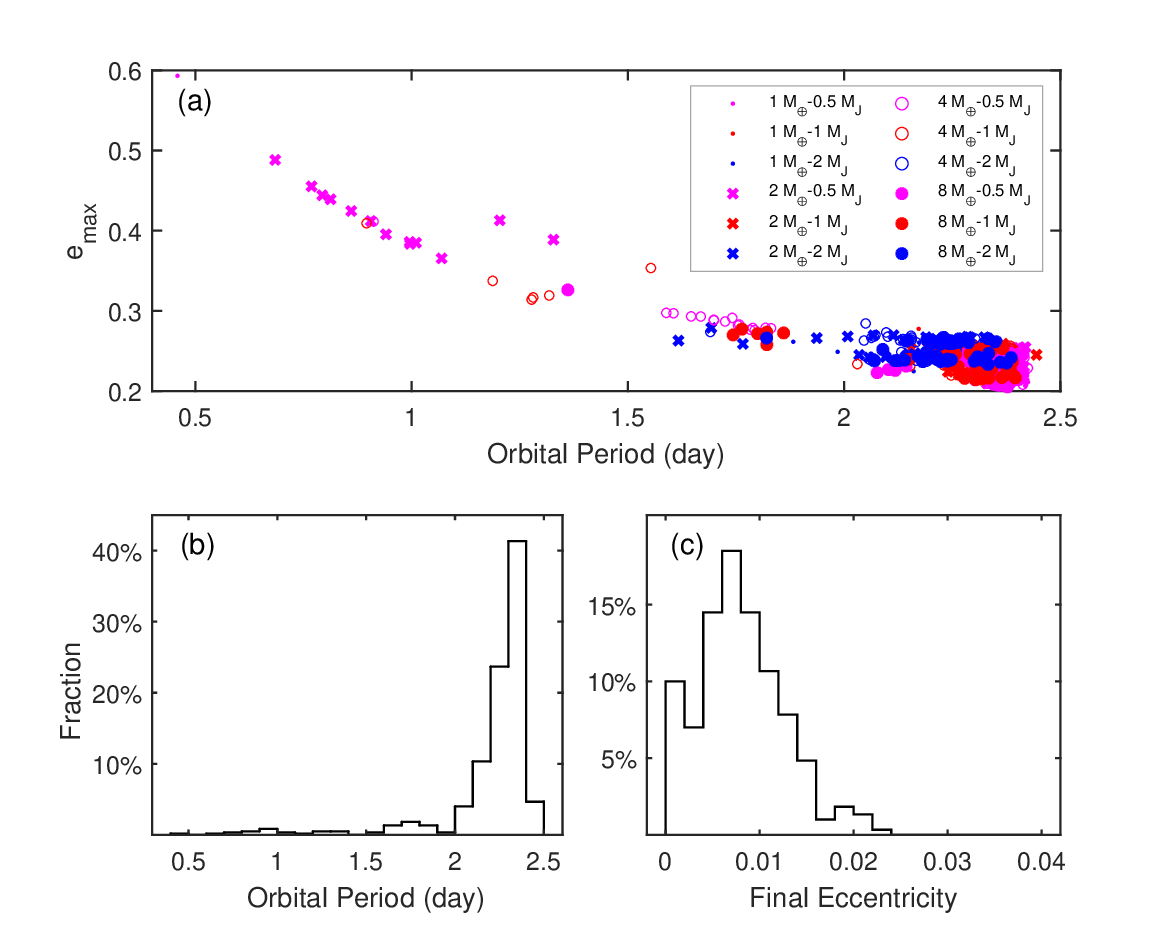}
    \caption{(a) Distribution of inner planets in 12 subgroups in group 1. The \textit{x}-axis represents the final orbital periods of the inner planets, while the \textit{y}-axis shows their maximum eccentricities during the evolution. Planetary masses of each subgroup are labeled by different shapes and colors. Different colors display different masses of the outer planets, and different shapes mean different masses of the inner planets. (b) The distribution of the final orbital periods of the inner planets. (c) The distribution of the final eccentricities of the inner planets.}
    \label{fig:4}
\end{figure*}

In this group, we perform 600 runs to examine how planetary mass affects the formation of USP planets. The evolution of typical Case 1 suggests that the formation of USP planets is closely related to eccentricity excitation. Panel (a) of Figure \ref{fig:4} shows the distribution of orbital periods and maximum eccentricities attained by planet b during its evolution. Different colors represent subgroups with different masses of the outer planets. Different shapes display subgroups with different masses of inner planets. From Panel (a), we can see that the USP planets are formed only if the eccentricity of it is excited to greater than 0.4.
Panel (b) of Figure \ref{fig:4} illustrates that 84.2\% of inner planets remain in the orbital period range of [2.0, 2.5] days, keeping the planet pair in a near 2:1 MMR, similar to the evolution process in case 2 as shown in Panels (c) and (d) of Figure \ref{fig:3}. Additionally, about 6.8\% of the simulations produce planetary configurations where the orbital period of the inner planet falls between 1.0 and 2.0 days. In these cases, eccentricity excitation occurs, but most eccentricities remain below 0.3, resulting in moderate orbital decay. USP planets with orbital periods less than 1.0 days emerge in only 1.7\% of the runs. These cases, similar to the evolution shown in Panels (a) and (b) of Figure \ref{fig:3}, exhibit eccentricity excitations exceeding 0.38, leading to rapid orbital decay below 1.0 days. The orbital crossing occurs in the remaining 7.3\% of runs. With the tidal effects, the eccentricities of most inner planets are damped to below 0.025 ultimately, as shown in Panel (c) of Figure \ref{fig:4}.

\begin{figure*}
    \centering
    \includegraphics[width=1\linewidth]{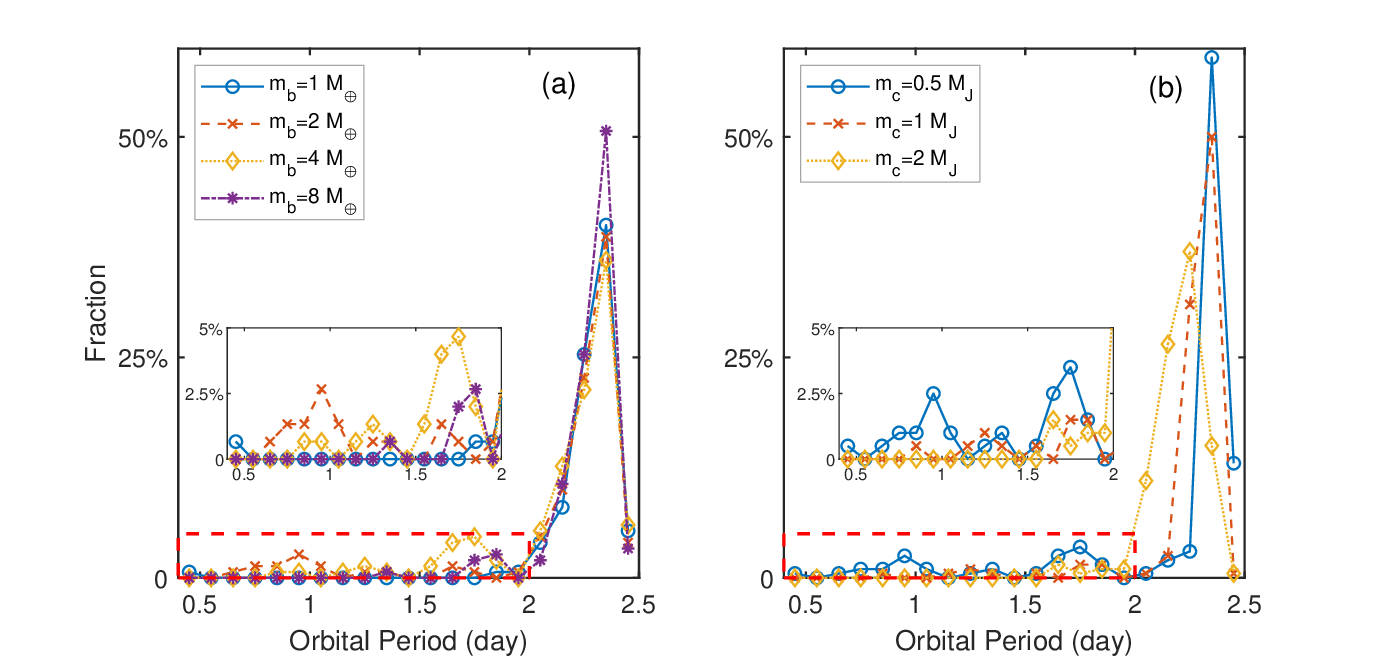}
    \caption{(a) Distribution of final orbital periods of inner planets with $M_{b}$ of 1, 2, 4, and 8 $M_\oplus$. (b) Distribution of final orbital periods of inner planets with $M_{c}$ of 0.5, 1, and 2 $M_{\rm J}$.}
    \label{fig:5}
\end{figure*}

Panel (a) of Figure \ref{fig:5} shows the results for fixed masses of the inner planets while varying the masses of the outer planets. As the size of the inner planet increases, its eccentricity is less likely to be excited to a higher value (as estimated by equation (\ref{circ})), leading to more planets remaining in the range of [2, 2.5] days. Most of the eccentricities of the inner planet with 8 $M_\oplus$ cannot be excited to greater than 0.3, which limits significant orbital decay. USP predominantly forms in the system where the inner planet is around 2 $M_\oplus$.
If the mass of the inner planet drops below 1 $M_\oplus$, its eccentricity can be excited beyond 0.6, often resulting in orbital crossings between planets. Therefore, as the mass of the inner planet decreases, its eccentricity is more likely to be excited. When the mass of the inner planet is greater than 1 $M_{\oplus}$, the eccentricities of the inner low-mass planets can be excited within the range of [0.3, 0.6], creating favorable conditions for the formation of USP planets. Figure \ref{fig:5} (b) shows that the distribution of the final orbital period of the inner planet varies with the mass of the outer planet. We can obtain the result that with the decrease in the mass of the outer planet, the fraction of USP increases, as shown by the blue line in this figure. This trend is more obvious in the zoomed-in inset of Panel (b). However, there are still a number of inner planets remaining in the range of [2.0, 2.5] days, and as the mass of the outer planet decreases, more planets settle closer to 2.5 days, suggesting that more massive outer planets tend to drive the planet pair further away from the MMRs.

\begin{figure*}
    \centering
    \includegraphics[width=1.0\linewidth]{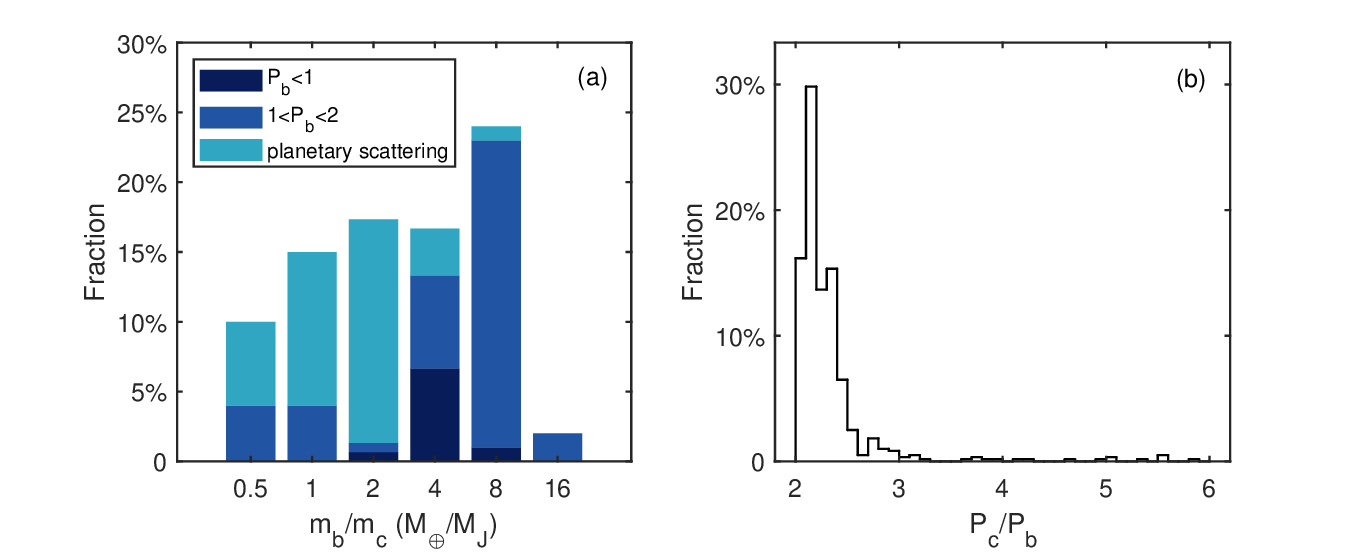}
    \caption{Panel (a) shows that the fraction of inner planets with different final orbital configurations varies with the mass ratios of the two planets. Different colors of bars show the probability of $P_{b}<1$ day, 1 d $<P_{b}<2$ days, and orbital crossing happened between two planets. The remaining portion of each bar represents the fraction of systems with inner planets in the range of 2-2.5 days. The mass ratios are in the range of 0.5-16 $M_{\oplus}/M_{\rm J}$. Panel (b) shows the distribution of period ratios of all the 600 runs in group 1.}
    \label{fig:8}
\end{figure*}

Panel (a) of Figure \ref{fig:8} illustrates how the fraction of different orbital configurations varies with the mass ratio of the two planets. The dark blue, royal blue, and dark cyan bars represent the planetary configurations with the orbital period of planet b less than 1 day, in the range of 1-2 days, and the configurations with orbital crossing, respectively. The remaining portion of each bar represents the fraction of systems with inner planets in the range of 2-2.5 days. For a mass ratio of 4 $M_{\oplus}/M_{\rm J}$, 10 cases out of 150 runs with USP planets formed, meaning approximately 6.7\% systems form USP planets. The highest probability of forming SP planets with orbital periods between 1 and 2 days is about 22\%,  which occurs when the mass ratio is 8 $M_{\oplus}/M_{\rm J}$. 6-16\% of cases with orbital crossing occur when the mass ratios of planets are in the range of 0.5-2 $M_{\oplus}/M_{\rm J}$. SP planets are more likely to form in systems with mass ratios exceeding 16 $M_{\oplus}/M_{\rm J}$.

Through the relationship between the orbital period evolution of planet b and the mass ratio of the two planets, we find that USP planets are most likely to form with $M_b/M_c \in (2-8)\ M_{\oplus}/M_{\rm J}$, which can maintain a stable orbit of less than 1 day. As the mass ratio increases, $M_b/M_c \geq 16\ M_{\oplus}/M_{\rm J}$, the smaller the disparity in mass between the two planets, the weaker the eccentricity excitation by the outer companion, and the higher the likelihood that two planets remain in a configuration of near 2:1 MMR. Conversely, as the mass ratio decreases, the eccentricity of the inner low-mass planet has a higher chance of being excited beyond 0.5, increasing the probability of orbital crossing.

Through the results shown in panel (b) of Figure \ref{fig:8}, we can find that the period ratio of 600 runs in group 1 distributes from 2.0 to even larger than 5.0. It means most of the inner planets are SP planets, ranging in the period ratio of 2-2.6 and near 2:1 MMR, and a few systems form USP planets.

\subsubsection{Group 2: Planetary Eccentricity}

\begin{figure*}
    \centering
    \includegraphics[width=1\linewidth]{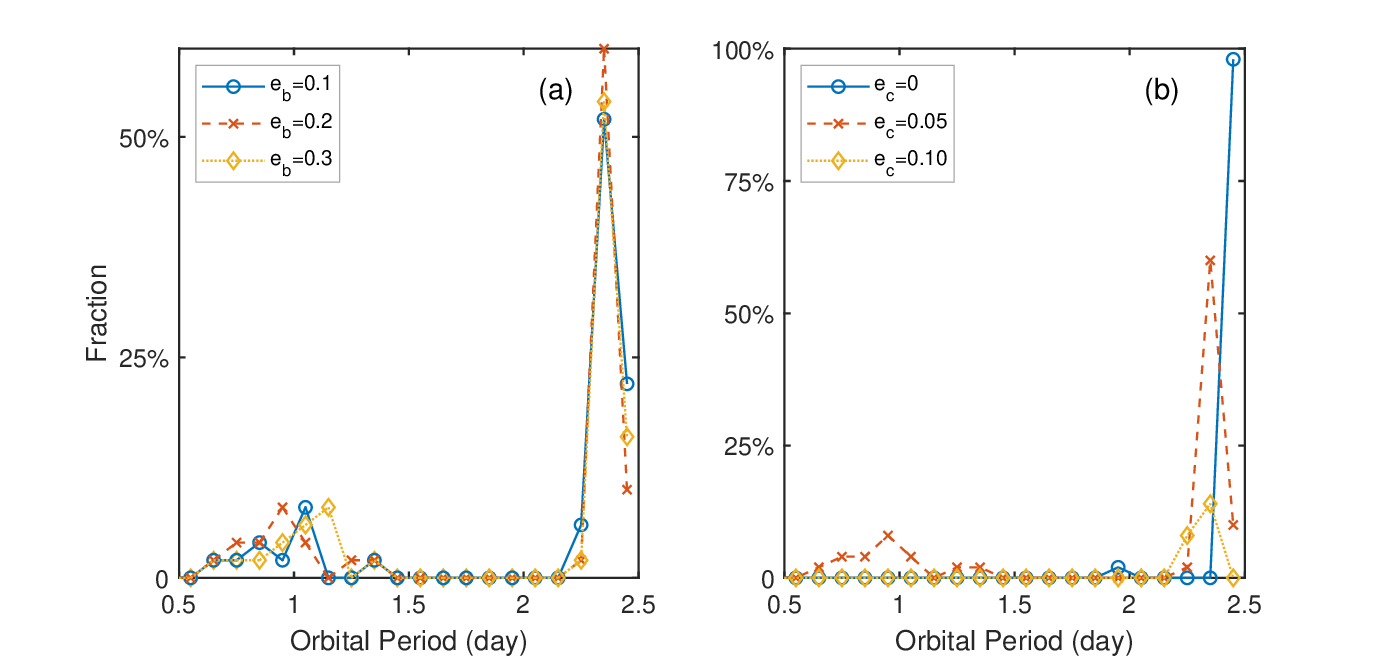}
    \caption{(a) Distribution of inner planets for different initial eccentricities of the inner planets, $e_{b0}$ = 0.1, 0.2, and 0.3, with the eccentricity of the outer companion fixed at 0.05. (b) Distribution of inner planets for different initial eccentricities of the outer companions, $e_{c0}$ = 0, 0.05, and 0.10, with the eccentricity of planet b set to 0.2.}
    \label{fig:9}
\end{figure*}

Since the eccentricity of the planet is related to the maximum excitation of the eccentricity, we examine the influence of initial eccentricities, $e_{b0}$ and $e_{c0}$, on the formation of USPs in group 2. As shown in Table \ref{mig}, during the orbital migration process, the eccentricity of planet b typically ranges from 0.1 to 0.3, while the eccentricity of planet c remains below 0.1 after they are trapped into MMR. Based on this, we set the initial eccentricities $e_{b0}$ = 0.1, 0.2, and 0.3, and $e_{c0}$ = 0, 0.05, and 0.10 in this group.

Panel (a) of Figure \ref{fig:9} presents the distribution of inner planets at the end of the simulations. In a small part of the system, the inner planets move closer to the central star, reaching orbital periods of 0.5-1.5 days, with their eccentricities excited to a maximum of 0.35-0.5. Another substantial fraction of systems evolve into a near 2:1 MMR configuration, with the inner planet moving slightly inward. However, regardless of their initial eccentricities, the probability that an inner planet will become a USP remains similar, ranging from 20\% to 26\%.

Panel (b) of Figure \ref{fig:9} illustrates the distribution of the inner planets for different initial eccentricities of the outer companions. When the companion HJ starts with a nearly circular orbit, the planet pairs will keep in a near 2:1 MMR configuration. The eccentricities of the inner planets rarely exceed 0.2. In such cases, almost all of the inner planets are in the region with orbital periods larger than 2.3 days. However, a significant fraction of inner planets collide with the outer HJ or are scattered out of the systems when the outer companion is in eccentric orbits with $e_c$ = 0.1, as shown in the yellow line in Panel (b) of Figure \ref{fig:9}. Only 22\% of the systems can keep two planets in a stable configuration with $e_{c0}$ = 0.1. With 0$<e_c<$0.1, the inner planet has a higher chance of evolving into a USP planet under the perturbation of the outer giant planet. The inner planet is seldom influenced by the perturbation from the outer HJ, which has an eccentricity close to 0. However, when $e_c\approx 0.1$, the likelihood of orbital crossing increases significantly. Overall, USP planets are more likely to form in systems where the outer companion has a small but nonzero eccentricity.

\subsubsection{Group 3: the Tidal Dissipation Function $Q_p$}

\begin{figure*}
    \centering
    \includegraphics[width=1\linewidth]{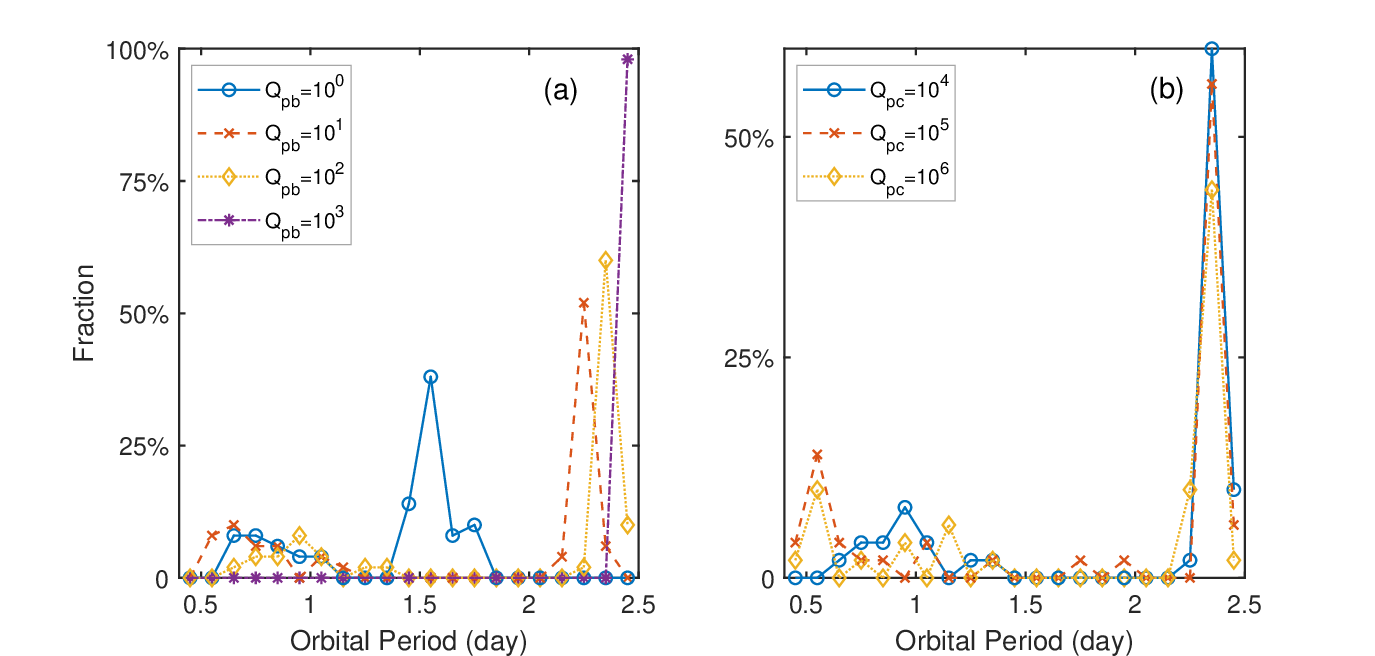}
    \caption{(a) Distribution of inner planets for different $Q_{pb}$ = $10^{0}$, $10^{1}$, $10^{2}$ and $10^{3}$, with $Q_{pc}$ fixed at $10^{4}$. (b) Distribution of inner planets for different $Q_{pc}$ = $10^{4}$, $10^{5}$ and $10^{6}$, with $Q_{pb}$ set to $10^{2}$.}
    \label{fig:10}
\end{figure*}

From equation (\ref{tidal}), we can find that the tidal circulation process is strongly influenced by the tidal dissipation function $Q_p$. In order to investigate the influence of the $Q$-value on the final configuration of the system, we consider the cases with a 2 $M_{\oplus}$ inner planet and a 0.5 $M_{\rm J}$ outer companion in the systems. The $Q_{pb}$ varies between $10^{0}$, $10^{1}$, $10^{2}$, and $10^{3}$, while for the outer companions, $Q_{pc}$ is set to $10^{4}$, $10^{5}$, and $10^{6}$.

\begin{figure*}
    \centering
    \includegraphics[width=0.8\linewidth]{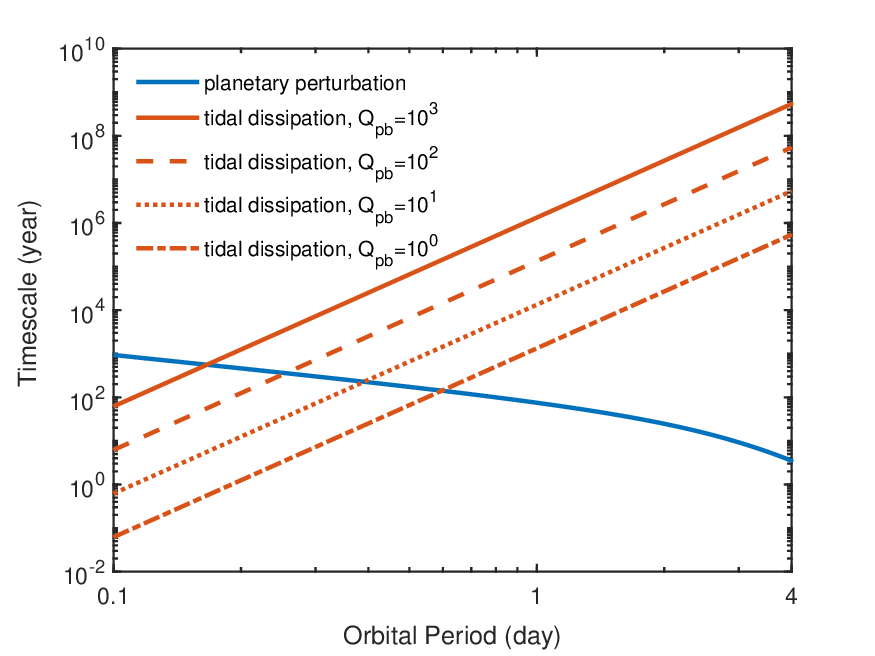}
    \caption{Timescales of secular perturbation from the outer giant and tidal dissipation of inner planet b in a system with $M_{b}=2\ M_{\oplus}$ and $M_{c}=0.5\ M_{\rm J}$, and the orbital period of hot Jupiter c is fixed at 5 days.}
    \label{fig:11}
\end{figure*}

Panel (a) of Figure \ref{fig:10} shows the orbital period distribution of the inner planet for different $Q_{pb}$. The timescales of secular perturbations and tidal effects on the inner planets when $Q_{pb}= 10^{0}, 10^{1}, 10^{2}$, and $10^{3}$ are illustrated in Figure \ref{fig:11}. The timescale of secular perturbation can be expressed as $\tau_{\rm per}=2\pi/(g_{1}-g_{2})$, where the $g_{1}$ and $g_{2}$ are two eigenfrequencies \citep{1999ssd..book.....M,2003ApJ...598.1290Z}. When the orbital period of the inner planet is near 2.5 days, the tidal circulation timescale is longer than the secular perturbation timescale. This allows the eccentricities of the inner planets to have an opportunity above 0.5, which is a key process in forming a USP planet. When the terrestrial planet has $Q_{pb} \leq 10^{2}$, the tidal effect from the host star is rapid enough to make the inner planets out of MMR, driving it inward. Once the inner planet enters the region below 1.5 days, the timescales for tidal damping of eccentricity and eccentricity excitation through secular perturbation become comparable. As a result, the eccentricity can still undergo small amplitude eccentricity excitations during the eccentricity damping by the tidal effects. This process leads to a continuous decrease in the planet’s semimajor axis, increasing the likelihood of it becoming a USP planet. For $Q_p\leq1$, the possibility that an SP planet forms with 1.3 days $<P<$ 2.0 days increases. With an increase in $Q_{pb}$, the inner planets have higher opportunities to stay closer to their outer companions, as shown in Panel (a) of Figure \ref{fig:10}. The peak position of SP planets shifts outward as the $Q_{pb}$ increases.

Figure \ref{fig:10} (b) presents the distribution of the orbital period of the inner planet for different $Q_{pc}$. 26$\%$ and 18$\%$ systems form USP planets with $Q_{pc} \sim 10^{5}-10^{6}$, among which 14$\%$ and 12$\%$ of the USP planets form within 0.7 days. The fraction of USP planets is much larger than that of systems with $Q_{pc} \sim 10^{4}$, where efficient tidal damping suppresses eccentricity growth in the outer companion. A higher $Q$-value of the outer giant extends its tidal dissipation timescale, allowing it to retain its eccentricity longer. This, in turn, enhances the perturbations acting on the inner planet, increasing the likelihood that its eccentricity is excited to higher values. As demonstrated in Section 3.2.2, the formation of USP planets is less likely when the eccentricity of the outer giant is low. Therefore, terrestrial planets have higher probabilities of becoming a USP planet with a large $Q_p$ of the outer massive planet. When $Q_{pc} \sim 10^{5}-10^{6}$, the formation probability of USP planets is similar, suggesting that for a $Q$-value comparable to that of Jupiter in the Solar system, the formation probability of a USP planet is not significantly different.

\section{CONCLUSION AND DISCUSSION}

In this work, we conducted numerical simulations to investigate the formation of USP planets in multiplanet systems under the influence of outer massive companions. In two-planet systems, the final planetary configurations are primarily shaped by key factors such as the mass ratio between the two planets, their eccentricities, and the efficiency of tidal dissipation. The main findings of our study are summarized as follows:

(a) Three distinct planetary configurations emerge due to the secular perturbation of an outer massive companion: (1) an HJ with an inner USP companion, (2) an HJ with an inner SP planet, and (3) a lonely HJ. Whether a USP forms depends on the competition between the eccentricity excitation and the damping process. Our simulations show that approximately 6.7\% of two-planet systems form USP planets. Under such a formation scenario, the period ratio between two planets is greater than 2.0 and extends up to 11.0. USP planets form when the eccentricity of the inner planet is excited beyond 0.4.

(b) The likelihood of USP planet formation is related to the mass of planets and the mass ratio between two adjacent planets. Inner planets with masses higher than 8 $M_\oplus$ have lower eccentricity excitation, leading to less opportunity to be USP planets. While a too-low mass of the inner planets causes strong eccentricity excitation, resulting in orbital crossing, which also leads to a reduced probability of USP formation, USP planets mainly form when the mass of the inner planet is 2 $M_\oplus$. USP planets form when the mass ratio $M_{\rm inner}/M_{\rm outer}$ falls within the range [2, 8] $M_\oplus/M_{\rm J}$, with the highest formation probability occurring in a ratio of $M_{\rm inner}/M_{\rm outer}$=4 $M_\oplus/M_{\rm J}$. With the increase of the mass ratio (i.e., the two planets become more comparable in mass), the inner low-mass planet is more likely to become an SP planet due to the weaker excitation of its eccentricity. Conversely, with small mass ratios between two planets, the eccentricity of the inner planet is more likely to be excited beyond 0.6, increasing the probability of orbital crossing and leaving behind a lonely HJ in the system.

(c) The eccentricity of the outer planetary companion plays a more significant role in shaping the final planetary configuration than the eccentricity of the inner planet. USP planets are more likely to form with $0<e_{\rm outer}<0.1$. If the outer massive planet is in a nearly circular orbit, the two planets tend to be stable in a near MMR configuration. However, when $e_{\rm outer}>0.1$, the inner planet struggles to maintain stability, making orbital crossing more likely and often leaving a lonely HJ in the system.

(d) USP planets tend to be formed in the system when the tidal circularization timescale is comparable to the secular perturbation timescale. This balance leads to multiple times of eccentricity excitations, which is a key factor resulting in a significant inward movement of the inner planet. In this study, USP planets are more efficiently formed when the tidal dissipation function of the inner planet is within the range $1.0<Q_{pb}<100$, similar to the values estimated for terrestrial planets in the solar system \citep{1999ssd..book.....M}. For outer HJs, a tidal dissipation function in the range of $10^5<Q_{p}<10^6$ can help maintain moderate eccentricity over long timescales, enhancing the likelihood of USP planet formation.

The scenario in this work is applicable for the formation of the planetary configuration of a USP planet or an SP planet with an outer massive planet near 10 days. Through confirmed planetary systems, eight systems contain an HJ or similar outer companion with an inner low-mass companion: TOI-1130, WASP-132, Kepler-730, Kepler-975, TOI-1408, WASP-47, WASP-84, and TOI-5398. In these systems, the period ratios between the massive planet and its inner low-mass companion range from 2.04 to 7.05. 60\% of the period ratio is less than 2.6, and in a near 2:1 MMR configuration, which is lower than our statistical results, as shown in Figure \ref{fig:8}. The large period ratios arise primarily from tidal effects on the innermost planet, which drive it closer to the central star. However, there are still other effects, such as the mass loss process \citep{2023AJ....165..174W} can lead to the formation of a large period ratio of planet pairs, which may increase the possibility that planet pairs escape from 2:1 MMR and form high period ratio configurations. The large period ratios arise primarily from tidal effects on the innermost planet, which drive it closer to the central star. Among the eight observed systems, most of the mass ratios between the inner planet and the outer massive planet are greater than 9 and in configurations consisting of SP planets accompanied by massive outer companions, which is consistent with the results obtained from our simulation model. Moreover, the eccentricities of the outer massive planets in these systems are generally close to 0, except for TOI-5398 b, which is another reason for the formation of SP planets in these systems. In the TOI-5398 system, although planet b has an eccentricity greater than 0.1, the relatively high mass ratio between the two planets may prevent the inner planet's eccentricity from being excited beyond 0.4. As a result, a USP planet is unlikely to form in such a system. Among the eight systems mentioned above, WASP-47 is the only confirmed system that contains both an HJ and a USP planet. The mass ratio between the two planets in WASP-47 is approximately 5.9 $M_{\oplus}/M_{\rm J}$, which falls within the range where USP planets are most likely to form around HJs. According to the formation scenario proposed here, the probability of a USP planet forming in a system with an outer massive planet among all the formed systems, including lonely HJs, is less than 10\%. This explains why most confirmed planetary systems feature an HJ with an inner SP planet companion or a lonely HJ. Future studies incorporating a broader range of initial conditions, additional dynamical effects, and observational constraints will be essential to further refine our understanding of USP planet formation and evolution.

\section{ACKNOWLEDGEMENT}
This work is supported by the National Natural Science Foundation of China (grant Nos. 12473076, 12033010, 12111530175, and 11873097), the Natural Science Foundation of Jiangsu Province (grant No. BK20221563), the B-type Strategic Priority Program of the Chinese Academy of Sciences (grant No. XDB41000000), the Foreign Expert Project (grant No. S20240145), the China Manned Space Project No. CMS-CSST-2021-B09 and No. CMS-CSST-2025-A16, Youth Innovation Promotion Association and the Foundation of Minor Planets of Purple Mountain Observatory.
\bibliography{Zhu20250822}{}

\end{CJK*}
\end{document}